\documentclass[review]{elsarticle}
\usepackage{hyperref}
\usepackage{verbatim} 
\usepackage{epsfig}
\usepackage{amssymb} 
\usepackage{mathtools}
\usepackage{tikz}        
\usepackage{etoolbox}
\usepackage{ulem}         
\usepackage[caption=false]{subfig}
\usepackage{textcomp}
\usepackage{float}
\usepackage{caption} 
\usepackage{booktabs}
\restylefloat{table}

\begin{document}

\begin{frontmatter}

\title{Improving competitive evacuations with a vestibule structure designed 
from panel-like obstacles}

\author[1]{I.M. Sticco}

\author[2]{G.A. Frank}

\author[1,3]{C.O. Dorso}

\address[1]{Departamento de F\'\i sica, Facultad de Ciencias Exactas y 
Naturales, Universidad de Buenos Aires,  Pabell\'on I, Ciudad Universitaria, 
1428 Buenos Aires, Argentina.}

\address[2]{Unidad de Investigaci\'on y Desarrollo de las Ingenier\'\i as, 
Universidad Tecnol\'ogica Nacional, Facultad Regional Buenos Aires, Av. Medrano 
951, 1179 Buenos Aires, Argentina.}

\address[3]{Instituto de F\'\i sica de Buenos Aires, Pabell\'on I, Ciudad 
Universitaria, 1428 Buenos Aires, Argentina.}

\begin{abstract}

It has been shown that placing an obstacle in front of an exit door has proven 
to be a successful method to improve pedestrian evacuations. In this work, we 
will focus on the space limited by the exit and the obstacles (\textit{i.e.} 
the vestibule structure). We analyzed two different types of 
vestibules: the two-entry vestibule (which consists of a single panel-like 
obstacle) and 
the three-entry vestibule (which consists of two panel-like obstacles). In the 
former, we studied the effects of varying the walls' friction coefficient 
$\kappa_w $ and the distance from the obstacle to the exit door $d$. In the 
latter, we varied the space between the two panels ($gap$). We found that the 
three above mentioned parameters control the vestibule's density, which 
subsequently affects the evacuation flow (fundamental diagram). We have also 
found that reducing the distance $d$ or increasing the friction facilitates the 
formation of blocking clusters at the vestibule entries, and hence, diminishes 
the density. If the density is too large or too low, the evacuation flow is 
suboptimal, whereas if the density is around 2\,p/m$^{2}$, the flow is 
maximized. Our most important result is that the density (and therefore the 
evacuation flow) can be precisely controlled by $\kappa_w$, $d$, and the $gap$. 
Moreover, the three-entry vestibule produced the highest evacuation flow for 
specific configurations of the gap and the distance from the panels to the 
door.\\

\end{abstract}


\end{frontmatter}


\section{Introduction}

The counterintuitive effect that an obstacle in front of a door can improve 
the performance of pedestrian evacuation is receiving more attention every year. 
 This novel assessment was suggested for the first time in the same paper where 
the escape panic version of the social force model was 
introduced~\cite{helbing2000simulating}. From that time, many studies have been 
conducted to exploit this phenomenon at its most.\\

Escobar \& De la Rosa performed numerical simulations using the escape panic 
version of the social force model to simulate pedestrian 
evacuations~\cite{escobar2003architectural}. They tested the performance of  
several architectural layouts in terms of the evacuation flow. They reported 
the existence of the ``waiting room effect''. The waiting room (also known as 
``vestibule'' in the architecture jargon~\cite{harris2006dictionary}) is the 
space in between the obstacles and the exit door, and its associated effect is 
the flow improvement whenever the inflow to this room does not exceed the 
outflow.\\

The first experiment of pedestrian evacuation using a column-like obstacle 
appeared in Ref.~\cite{helbing2005self} (to our knowledge). Although the 
number of participants was small ($N=20$), the results are qualitatively 
consistent with the previous numerical simulations. Frank \& Dorso performed 
numerical simulations of evacuations using panel-like 
obstacles~\cite{frank2011room}. They 
show that, under certain conditions, the panel-like 
obstacles may exceed the performance of pillar-like obstacles. They also 
report the ``clever is not always better'' effect, meaning that the 
overall evacuation may be improved if the agents follow a non-strategic plan in 
order to avoid the obstacles. \\

Although most of the investigations focused on the pillar-like 
obstacle~\cite{zuriguel2020contact,garcimartin2018redefining,shi2019examining,
liu2016controlled,shiwakoti2013enhancing}, the panel-like obstacle appears to 
perform better in the evacuation process, and thus, has brought the interest in 
the last years~\cite{frank2011room,zhao2017optimal,li2019emergency}. The first 
laboratory experiment using a panel-like obstacle was performed by Haghani \& 
Savri~\cite{haghani2019simulating}. Under certain experimental conditions, they 
found that the panel improves the 
evacuation if the doors are narrow enough. \\

Ref.~\cite{zhao2017optimal}, performed a numerical optimization for the   
dimensions and location of panel-like and pillar-like obstacles. They claim 
that the former perform better than the latter because of the capability 
to reduce the egress time and the pressure among 
pedestrians. They emphasize that these conclusions hold even for a wide set of 
parameters. A more recent investigation carried out by the same group, 
experimentally confirms that panels yield better performance than the 
pillars~\cite{zhao2020experimental}.\\

Although most of the researches state that the panel-like obstacle improves the 
evacuation, other novel results challenge this 
statement~\cite{ding2020evacuation}. For instance, a recent study 
explores the limitations of the panel-like obstacle when the pedestrians face a 
limited-visibility situation~\cite{li2019emergency}.\\

For a complete review on the obstacle effects in many diverse systems, 
(including cellular 
automata~\cite{kirchner2003friction,yanagisawa2010study,matsuoka2015effects}), 
see Ref.~\cite{shiwakoti2019review}. It should be mentioned that the interest in 
the obstacle's effect went beyond the pedestrian dynamics field since it has 
been further explored in granular 
media~\cite{zuriguel2011silo,lozano2012flow,alonso2012bottlenecks} and 
bottleneck egress with 
animals~\cite{shiwakoti2009enhancing,garcimartin2015flow,zuriguel2016effect}. \\

One of the most common ways to quantify the capacity of a pedestrian facility 
is by means of the fundamental diagram. That is, the flow-density 
relation (or velocity-density relation)~\cite{seyfried2005fundamental}. The 
fundamental diagram exhibits two different regimens: the 
free-flow regime, in which the flow increases as the density increases, and the 
congested regime, in which the flow decreases as the density increases. The 
fundamental diagram has been studied in many contexts ranging from 
laboratory 
conditions~\cite{seyfried2005fundamental,ren2021flows,cao2017fundamental} to 
real-life situations~\cite{helbing2007dynamics,lohner2018fundamental}. A more 
detailed insight into this topic can be found 
in Refs.~\cite{kretz2019overview,vanumu2017fundamental,cao2018investigation}.\\

The fundamental diagram at bottlenecks is still a subject of debate 
due to the lack of consensus on the expected regimes. Kuperman et al., for 
instance, performed a controlled 
experiment where they measured the egress flow as a function of the pedestrian 
density close to the exit door~\cite{nicolas2017pedestrian}. Contrary to 
expected results, they report an increasing flow even for density values as 
high as $\rho=9\,$p/m$^{2}$.\\

Ref.~\cite{bernardini2016towards}, obtained the fundamental diagram of a 
real-life earthquake evacuation. Although the 
density values did not exceed $\rho=6\,$p/m$^{2}$, they found a 
non-monotonically flow increment versus density. On the other hand, 
the laboratory experiments appearing in Ref.~\cite{daamen2005first} exhibit a 
plateau in the flow after the capacity (the maximum flow) is achieved. 
Nevertheless, other laboratory experiments in bottleneck 
environments display the typical flow diminution that characterizes the 
congested regime~\cite{seyfried2009new,bukavcek2015experimental}. \\

The discrepancy across the literature results may be attributed to the 
differences in the door's width, the disparity in the participants' anxiety 
level and the variation between measurement areas (among other reasons). It 
should be mentioned, though, that the fundamental diagram is highly sensible to 
the flow measurement criteria as thoroughly analyzed in 
Ref.~\cite{seyfried2010enhanced}. Another consideration to take into account is 
that controlled experiments are not suitable for reproducing panic conditions 
since participants are not allowed to harm each other (due to safety reasons).\\

It is worth mentioning that the link between the flow-density relation and 
pedestrian evacuations in the presence of obstacles is almost 
unexplored~\cite{haworth2017density}. In this paper, we will bridge the 
gap between the fundamental diagram and the evacuations in the presence of 
panel-like obstacles. \\

\section{\label{background} Background}

\subsection{\label{sfm}The Social Force Model}

The social force model~\cite{helbing2000simulating} provides a necessary 
framework for simulating  the collective dynamics of pedestrians (\textit{i.e.} 
self-driven agents). The pedestrians are represented as simulated agents that 
follow an equation of motion involving  either ``socio-psychological'' forces 
and physical forces. The equation of motion for any agent $i$ of mass $m_i$ 
reads

\begin{equation}
 m_i\,\displaystyle\frac{d\mathbf{v}_i}{dt}=\mathbf{f}_d^{(i)}+
 \displaystyle\sum_{j=1}^N\mathbf{f}_s^{(ij)}+
 \displaystyle\sum_{j=1}^N\mathbf{f}_p^{(ij)}\label{newton_ec}
\end{equation}

\noindent where the subscript $j$ corresponds to any neighboring agent or
the walls. The three forces $\mathbf{f}_d$, $\mathbf{f}_s$ and $\mathbf{f}_p$
are different in nature. The desire force $\mathbf{f}_d$ represents the
acceleration of a pedestrian due to his/her own will.  The
social force $\mathbf{f}_s$, instead, describes the tendency of the  pedestrians
to stay away from each other (social distancing). The physical force 
$\mathbf{f}_p$  stands for both the sliding friction and the repulsive body 
force. \\

The pedestrians' own will is modeled by the desire force $\mathbf{f}_d$.  This
force stands for the acceleration required to move  at the
desired walking speed $v_d$. The parameter $\tau$ reflects the reaction time. 
Thus, the desire force is modeled as follows

\begin{equation}
\mathbf{f}_d^{(i)}=m\,\displaystyle\frac{v_d^{(i)}\,
\hat{\mathbf{e}}_d^{(i)}(t)-
 \mathbf{v}^{(i)}(t)}{\tau}
\end{equation}

\noindent where $\hat{\mathbf{e}}(t)$ represents the unit vector pointing to the
target position and $\mathbf{v}(t)$ stands for the agent velocity at time $t$.\\

The tendency of any individual to preserve his/her personal space is
accomplished by the social force $\mathbf{f}_s$. This force is expected to
prevent the agents from getting too close to each other (or to the 
walls) in any environment. The model for this kind of  ``socio-psychological''
behavior is as follows

\begin{equation}
 \mathbf{f}_s^{(i)}=A\,e^{(R_{ij}-r_{ij})/B}\,\hat{\mathbf{n}}_{ij}
 \label{eqn_social}
\end{equation}

\noindent where $r_{ij}$ means the distance between the center of mass of the
agents $i$ and $j$, and $R_{ij}=R_i+R_j$ is the sum of the pedestrians
radius. The unit vector $\hat{\mathbf{n}}_{ij}$ points from pedestrian $j$ to
pedestrian $i$, meaning a repulsive interaction. The parameter $B$ is a 
characteristic scale that plays the role of a fall-off length within the social 
repulsion. At the same time, the parameter $A$ represents the intensity of the 
social repulsion. \\ 

The  expression for the physical force (the friction force plus the body 
force) has been inspired from the granular  matter 
field~\cite{risto1994density}. The mathematical expression reads as follows

\begin{equation}
 \mathbf{f}_p^{(ij)}=\kappa_t\,g(R_{ij}-r_{ij})\,
(\Delta\mathbf{v}^{(ij)}\cdot\hat{\mathbf{t}}_{ij})\,\hat{\mathbf{t}}_{ij}+
k_n\,g(R_{ij}-r_{ij})\,
\,\hat{\mathbf{n}}_{ij}\label{eqn_friction}
\end{equation}

\noindent where $g(R_{ij}-r_{ij})$ equals $R_{ij}-r_{ij}$ if $R_{ij}>r_{ij}$ and
vanishes otherwise. $\Delta\mathbf{v}^{(ij)}\cdot\hat{\mathbf{t}}_{ij}$
represents the relative tangential velocities of the sliding  bodies (or between
the individual and the walls).    \\

The sliding friction occurs in the tangential direction while the body force
occurs in the normal direction. Both are assumed to be linear with respect to
the net distance between contacting agents. The sliding friction is also
linearly related to the difference between the tangential velocities. The
coefficients $\kappa_t$ (for the sliding friction) and $k_n$ (for the  body 
force) are related to the contacting surfaces materials and the 
body stiffness, respectively. \\

The model parameter values were chosen to be the same as the 
best-fitting parameters reported in the recent study from 
Ref.~\cite{sticco2021social} These values were obtained by fitting the 
model to a real-life event that resemblances an emergency 
evacuation. The parameter values are: $A=2000\,$N, $B=0.08\,$m, 
$\kappa_t=3.05\times10^{5}\,$kg/(m.s), $k_n=3600\,$N/m, $\tau=0.5\,$s.\\

The friction force between an agent and a wall has the same mathematical 
expression as the friction between two agents. Nevertheless, we distinguish the 
friction coefficient between agents $\kappa_t$ and the wall friction 
coefficient $\kappa_w$. In this paper, we will fix the value of $\kappa_t$, but 
we will explore different values of $\kappa_w$.\\

\subsection{\label{bc} Blocking clusters}

A characteristic feature of pedestrian dynamics is the formation of clusters.
Clusters of pedestrians can  be defined as the set of individuals that for any
member of the group (say, $i$) there exists at least another member belonging to
the same group ($j$) in contact with the former.  Thus, we define a ``granular
cluster'' ($C_g$) following the mathematical formula given in
Ref.~\cite{strachan1997fragment}

\begin{equation}
C_g:P_i~\epsilon~ C_g \Leftrightarrow \exists~ j~\epsilon~C_g / r_{ij} 
< (R_i+R_j) \label{ec-cluster}
\end{equation}

where ($P_i$) indicate the \textit{ith} pedestrian and $R_i$ is his/her radius
(half of the shoulder width). That means, $C_g$ is a set of pedestrians that 
interact not only with the social force, but also with physical forces 
(\textit{i.e.} friction force and body force). A ``blocking cluster'' is 
defined as the minimal granular cluster which is closest to the door whose first 
 and last agents are in contact with the walls at both sides of the 
door ~\cite{parisi2005microscopic}. Previous studies have demonstrated that 
the blocking clusters play a crucial role in preventing pedestrians from getting 
through a door~\cite{parisi2005microscopic,sticco2017room,cornesmicroscopic}.

\section{Numerical simulations}

\begin{figure}[!htbp]
\centering
\subfloat[]{\includegraphics[trim=0 -0.18cm 0 0, width=0.49\columnwidth] 
{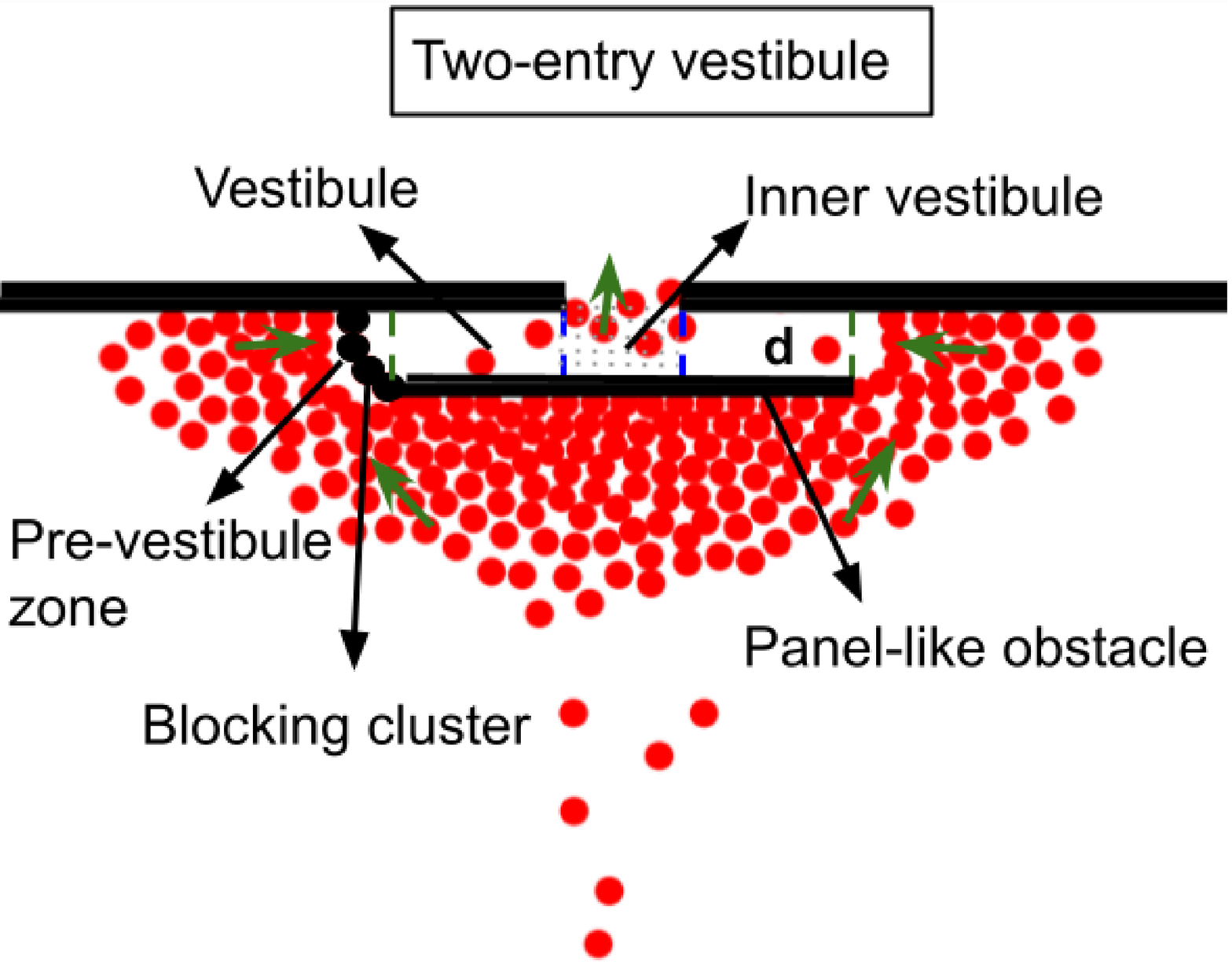}\label{two-entry_vestibule}}\ 
\subfloat[]{\includegraphics[width=0.49\columnwidth]
{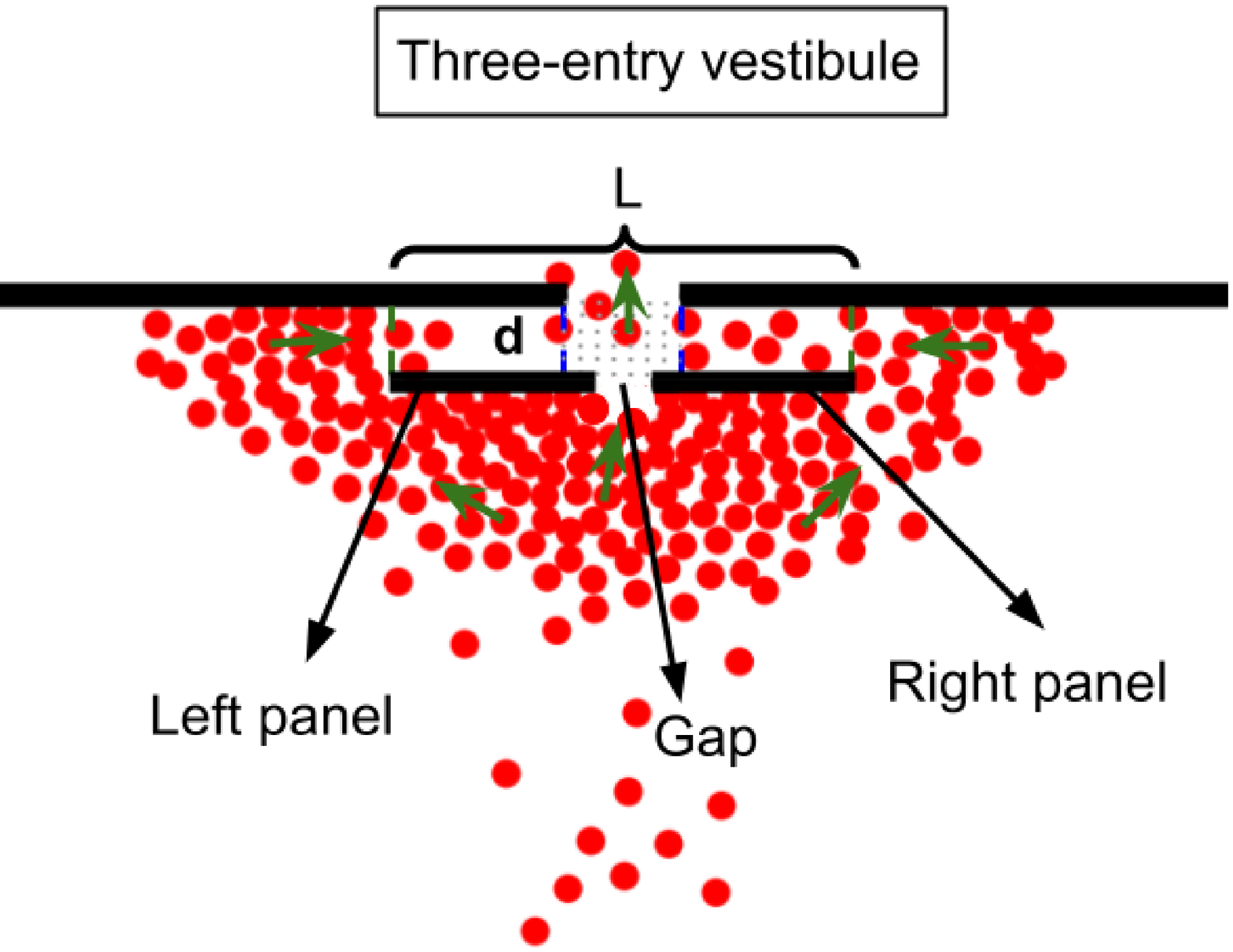}\label{three-entry_vestibule}}\\
\caption[width=0.49\columnwidth]{Visual representations of the numerical 
simulations, the red circles represent the agents. The space between the 
obstacle and the walls is the vestibule (delimited by green vertical dashed 
lines). The letter `d' on the vestibule denotes the vertical distance between 
the panels and the door. The ``inner vestibule'' is the dotted area enclosed by 
vertical blue lines (close to the exit door). The green arrows stand for the 
desired velocity. \textbf{(a)} Two-entry vestibule structure. The black 
agents located at the left entrance to the vestibule make up a blocking 
cluster in the pre-vestibule zone. \textbf{(b)} Three-entry vestibule 
structure.}
\label{rooms_layout}
\end{figure}

We carried out our investigation by performing numerical simulations of 
pedestrians evacuating a room in the presence of one or two obstacles, 
as it is shown in Fig.~\ref{rooms_layout}. We simulated a crowd of $N=200$ 
agents 
whose trajectories followed a classical Newton equation of motion. We used the 
escape panic version of the social force model for the interaction forces acting 
on the agents (see Section \ref{sfm} for the mathematical 
expressions). The model parameter values were chosen to be the same as the 
best-fitting parameters reported in the recent study from 
Ref.~\cite{sticco2021social}, the values were introduced in Section \ref{sfm}. 
\\ 

The mass of the agents was fixed at $m=80\,$kg, and the radius was set to 
$R=0.23\,$m 
according to the data from Ref.~\cite{littlefield2008metric}. The desired 
velocity (which is the parameter that controls the anxiety level) was varied in 
the interval $0.5\,$m/s\,$\leq v_d\leq6\,$m/s. The upper limit ($v_d=6\,$m/s) is 
a velocity high enough to represent a rush but low enough to be achieved by 
non-professional runners~\cite{morin2012mechanical}. 
Initially, the agents were placed in a $20$\,m\,$\times \, 20\,$m area with 
random positions and velocities. The direction of the desired velocity was 
such that the agents located outside the vestibule point to the vestibule 
through the nearest entry. 
Once they were inside the vestibule, the target became the exit door (see the 
green arrows in Fig.~\ref{rooms_layout}). \\

The exit door's size was $DS=1.84\,$m, which is equivalent to four agents' 
diameter in accordance to Ref.~\cite{sticco2021social}. The two-entry vestibule 
(Fig.~\ref{two-entry_vestibule}) has a single unidimensional panel of size $L=4 
\times DS= 7.36\,$m centered with respect to the door (in the same way as in 
Ref.~\cite{frank2011room}). Notice that it is possible to access the vestibule 
from the left or right sides (these are the two-entries).\\

The three-entry vestibule (Fig.~\ref{three-entry_vestibule}) has two panels of 
the same size. We call it ``three-entry'' because the gap between panels 
becomes the third entry (in addition to the left and right entries). The $gap$ 
is defined as the space between panels. The size of the panels was constrained 
by the gap between them. The distance between the ends of the panels was fixed 
at $L=7.36\,$m (see Fig.~\ref{three-entry_vestibule}).\\

We analyzed the effect of varying four parameters: the desired velocity $v_d$, 
the wall friction coefficient $\kappa_w$, the vertical distance between the 
obstacle and the door $d$ and, in the case of the three-entry vestibule, we 
varied the gap between panels ($gap$). We stress that the wall friction 
coefficient is also the panel's friction coefficient since, in this paper, 
panels are treated as walls. It should be noted that the three parameters are 
architectural features that could be easily handled by designers and 
contractors. \\

For each parameter configuration, we performed 30 evacuation processes that 
finished when 90\% of agents left the room. No re-enter of agents was allowed. 
We measured the mean density, the evacuation flow, and the probability of 
blocking clusters. The mean density was measured in the inner 
vestibule region. It was defined as the number of agents divided by the 
corresponding occupied area. The evacuation flow was defined as the the 
number of evacuated agents divided by the evacuation time. The blocking clusters 
are the set of pedestrians that block a 
door (see Section \ref{bc} for a formal definition). We 
measured the probability of attaining blocking clusters at the entries of the 
two-entry vestibule (at the pre-vestibule zone). The probability was defined as 
the fraction of time 
with blocking clusters over the total evacuation time. \\

The numerical simulations were carried out using LAMMPS, which is a molecular 
dynamics open-access software~\cite{plimpton1995fast}. The implementation also 
required customized modules developed in C\raisebox{1mm}{\tiny ++}. 
The integration of the agents' trajectory was calculated using the 
Velocity Verlet algorithm with a time-step $\Delta t=10^{-4}\,$s.\\

\subsection{Clarifications}

In order to keep this paper concise and clear, we will only report the results 
corresponding to the numerical simulations described above. Nevertheless, we 
want to mention that we performed numerical simulations with higher number of 
agents $N=400$ (instead of $N=200$) and we obtained similar qualitative 
results to the ones that will be reported in Section \ref{results}. We also 
tested the results' robustness by replacing the unidimensional panels with 
obstacles of width $w=0.12\,$m~\cite{frank2011room} and no significant 
differences were observed.\\

We measured the density in the inner vestibule and the density in the whole 
vestibule (see Fig.~\ref{rooms_layout}). In both cases, the mean values are 
qualitatively similar. This paper only reports the results corresponding to the 
inner vestibule density because the data exhibit less dispersion around the mean 
values.\\

For simplicity reasons, we will omit the units of friction coefficient 
$\kappa_w$. 
Bear in mind that its units are kg/(m.s). The gap between panels will be 
express in units of agents' diameter. In this context, we will use the letter 
`p' to refer to the agent diameter. For instance, $gap = 3\,$p means that the 
gap size is $gap = 3\times 0.46\,$m (since the agent diameter is $0.46\,$m). 
The distance between the panel and the door $d$ will be presented like in the 
following example: $d = 0.92\,$m ($2\,$p), where `$2\,$p' stands for two 
agents' diameter.\\

\section{\label{results} Results and discussions}

We present in this Section the main results of our investigation. We divided 
this Section into three parts. In the first part (Sec.~\ref{flow_vs_vd}), we 
show the evacuation flow as a function of the desired velocity for both the 
two-entry vestibule and the three-entry vestibule.\\

In the following two parts, we explore only the highest desired velocity 
$v_d=6\,$m/s because this desired velocity corresponds to an anxious crowd that 
could lead to a dangerous situation. In Section~\ref{two-entry vestibule} we 
present the results of the two-entry vestibule  and the effects of varying 
either the friction of the walls 
and the distance of the obstacle to the door. In Section~\ref{three-entry 
vestibule}, 
we explore the consequences of placing two panel-like obstacles in front of the 
door (\textit{i.e.} a three-entry vestibule) and the effects produced by 
varying the gap between the panels.\\

\subsection{\label{flow_vs_vd}Flow vs. desired velocity}

In this Section, we show results corresponding to both the two-entry vestibule 
and the three-entry vestibule. We will show the consequences on the flow after 
varying the desired velocity and the obstacle's size. The desired velocity is 
the parameter that controls the pedestrians' anxiety 
level. If the desired velocity is low (say, $v_d=1\,$m/s), it means that 
pedestrians are in a relaxed situation. On the other hand, a higher desired 
velocity (say, $v_d=6\,$m/s) means that pedestrians rush to exit the room. 
\\

\begin{figure}[!htbp]
\centering
\subfloat[]{\includegraphics[width=0.49\columnwidth]
{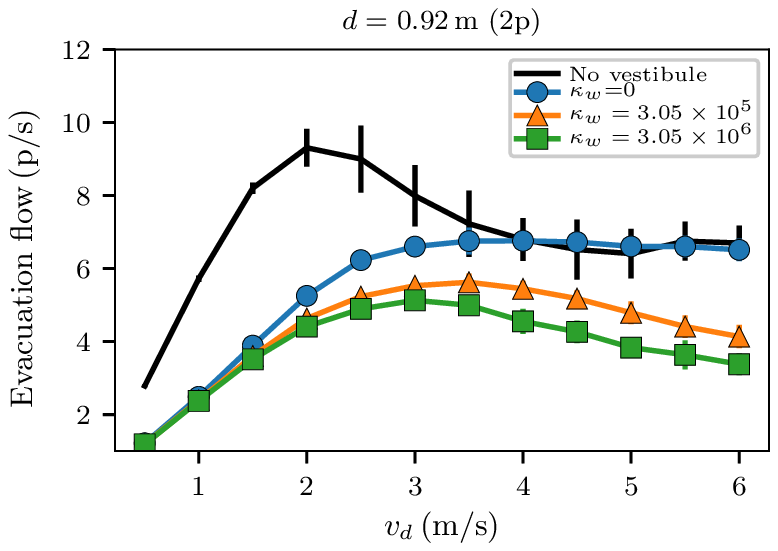}\label{low_vs_vd_d0.92}}\ 
\subfloat[]{\includegraphics[width=0.49\columnwidth]
{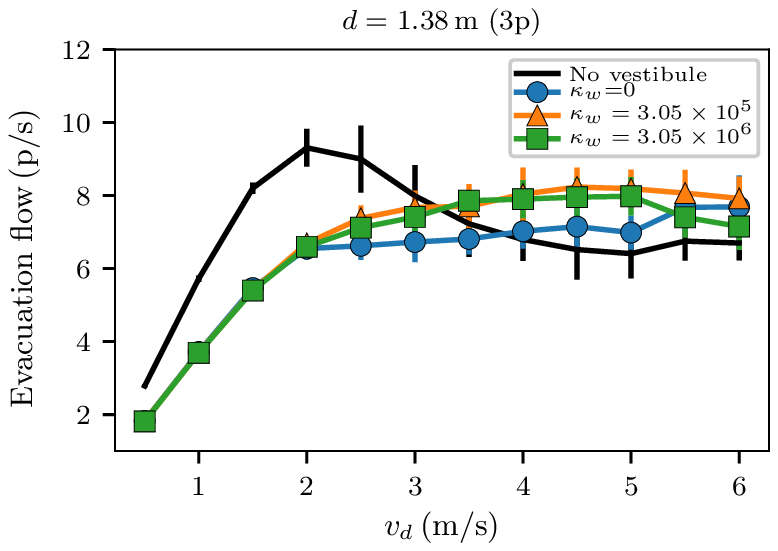}\label{flow_vs_vd_d1.38}}\\
\subfloat[]{\includegraphics[width=0.49\columnwidth]
{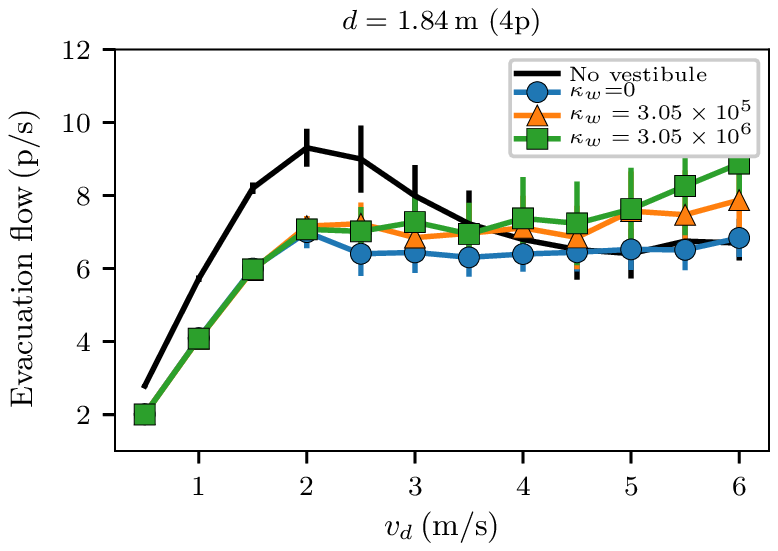}\label{flow_vs_vd_d1.84}}\ 
\subfloat[]{\includegraphics[width=0.49\columnwidth]
{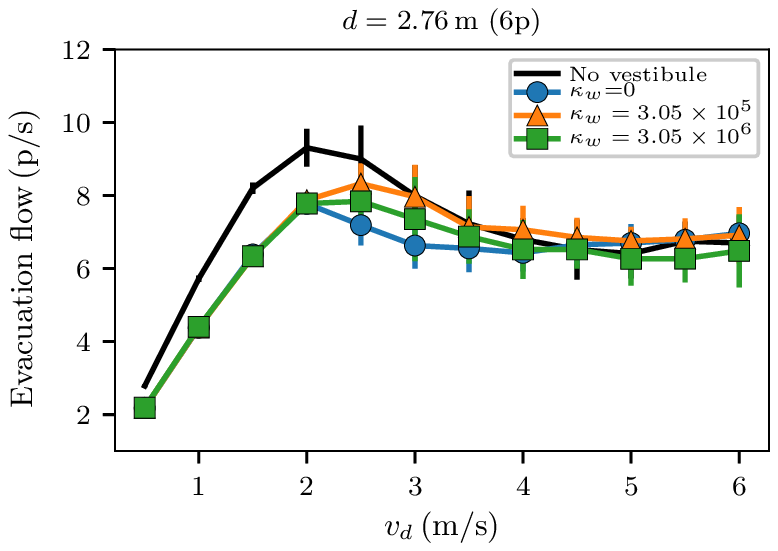}\label{flow_vs_vd_d2.76}}\\
\caption[width=0.47\columnwidth]{Evacuation flow as a function of the desired 
velocity for two-entry vestibules. The measurements are averaged over 30 
evacuation processes where the initial positions and velocities were set at 
random. The initial number of  pedestrians was $N=200$. See the plot's title 
for the corresponding distance from the obstacle to the door. See the legend 
for the wall friction coefficient value $\kappa_w$.}
\label{flow_vs_vd_1panel}
 \end{figure}

Fig.~\ref{flow_vs_vd_1panel} shows the evacuation flow as a function of the 
desired velocity for the two-entry vestibule. The flow is defined as $J=n/t_e$ 
where $n$ is the number of evacuated pedestrians and $t_e$ is the time it takes 
to evacuate those $n$ pedestrians. We show four ``subplots'', each of them 
corresponds to different distances between the panel and the exit door (see the 
plots' titles). Each curve is associated with a particular 
wall friction coefficient $\kappa_w$. We set $\kappa_w=$0 as a limiting 
case where no wall friction is present, $\kappa_w=3.05\times 10^{5}$ 
corresponding to the value obtained from the optimization of parameters in 
Ref.~\cite{sticco2021social} and $\kappa_w=3.05\times 10^{6}$ as the upper limit 
friction value explored in this research.\\ 

We explored four different conditions: the ``no-vestibule'' situation and the 
vestibule situation with three different wall friction values shown in the 
legends.\\

As a general feature, we observe that if the desired velocity is low 
($v_d<4\,$m/s \textit{i.e.} almost relaxed situation), the scenario without 
vestibule produces higher evacuation flows than the scenarios that includes it. 
Nevertheless, if the desired velocity is high enough ($v_d\ge4\,$m/s say, when 
the multitude is anxious and the risk of fatalities is present), the 
vestibule improves the evacuation performance for panels placed at $d=1.38\,$m 
and $d=1.84\,$m; see Figs. \ref{flow_vs_vd_d1.38} and \ref{flow_vs_vd_d1.84} in 
the range $v_d\ge4\,$m/s.\\

\begin{figure}[!htbp]
\centering
\subfloat[]{\includegraphics[width=0.49\columnwidth]
{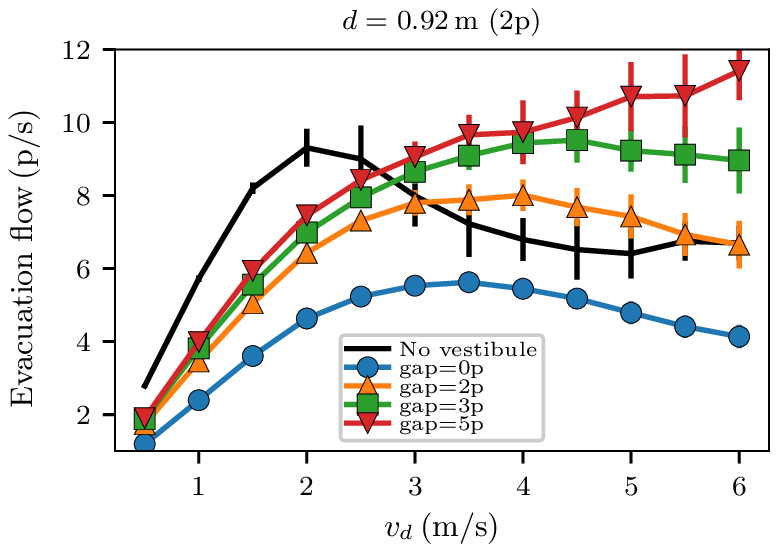}\label{flow_vs_vd_d0_92_2panels}}\ 
\subfloat[]{\includegraphics[width=0.49\columnwidth]
{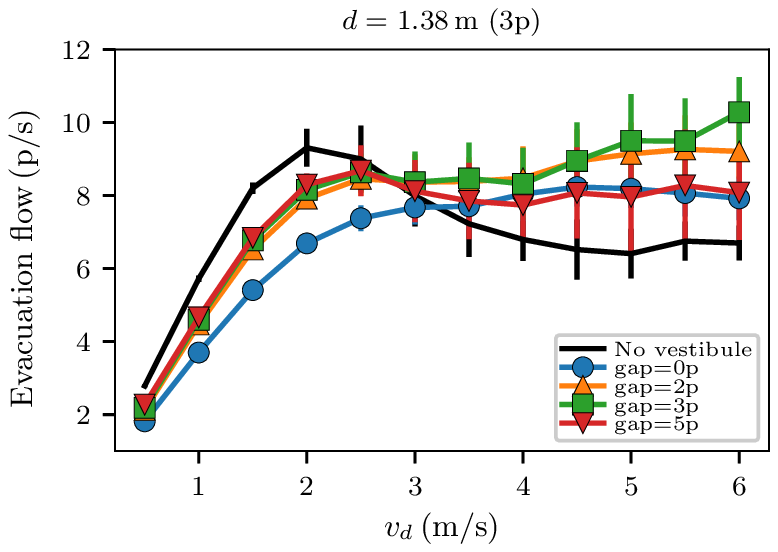}\label{flow_vs_vd_d1_38_2panels}}\\
\subfloat[]{\includegraphics[width=0.49\columnwidth]
{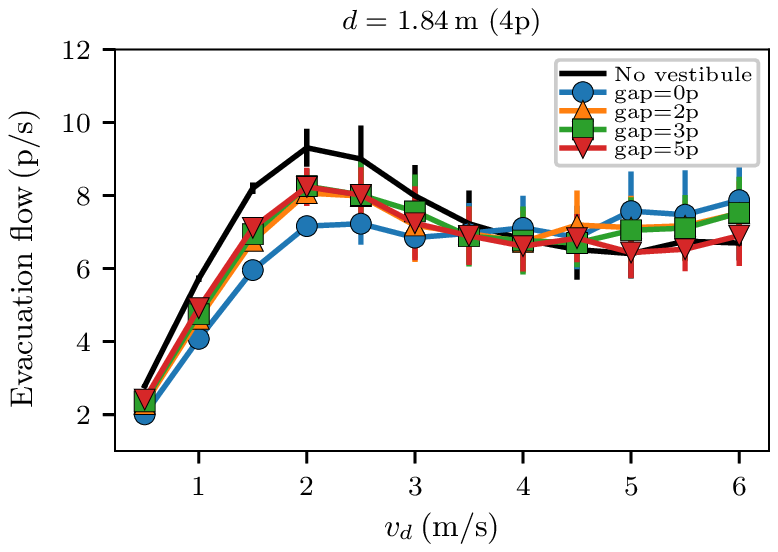}\label{flow_vs_vd_d1_84_2panels}}\
\caption[width=0.47\columnwidth]{Evacuation flow as a function of the desired 
velocity for three-entry vestibules. The data was averaged over 30 
evacuation processes. The initial conditions were similar to those in 
Fig.~\ref{flow_vs_vd_1panel}. The initial number of  pedestrians was $N=200$. 
See the plot's title for the corresponding distance from the obstacles to the 
door. See the legend for the gap between the obstacles.}
\label{flow_vs_vd_2panels}
\end{figure}

We measured the evacuation flow as a function of $v_d$ for the three-entry 
vestibule structure. Fig.~\ref{flow_vs_vd_2panels} shows the results 
corresponding to different distances between the panels and the exit door (see 
the plots' titles). In this case, we varied the gap between the panels as 
shown in the legends. \\

We may conclude, from this Section, that the vestibule improves the evacuation 
performance (in terms of the evacuation flow) if the desired velocity is high 
enough (and withing the explored range). This suggests that the vestibule 
structure may facilitate the egress of pedestrians if they feel a sudden urge to 
escape from a particular place (such as in an emergency evacuation). In the 
following two Sections, we set the desired velocity to $v_d=6\,$m/s. This 
value was chosen because it is high enough to ensures a high anxiety level but 
low enough to be a velocity achievable by non-professional runners. Bear in mind 
that similar results can be obtained for desired velocities in the range 
$v_d\ge4\,$m/s, where pressure conditions become dangerous.\\

\subsection{\label{two-entry vestibule}The two-entry vestibule}

In this Section, we present and discuss the room evacuation results of a 
vestibule with two entries (left and right entries). The room layout is 
illustrated in Fig.~\ref{two-entry_vestibule}.

\subsubsection{\label{Wall friction and distance effects}  Effects of the 
distance and the wall friction}

\begin{figure}[!htbp]
\centering
\subfloat[]{\includegraphics[width=0.49\columnwidth]
{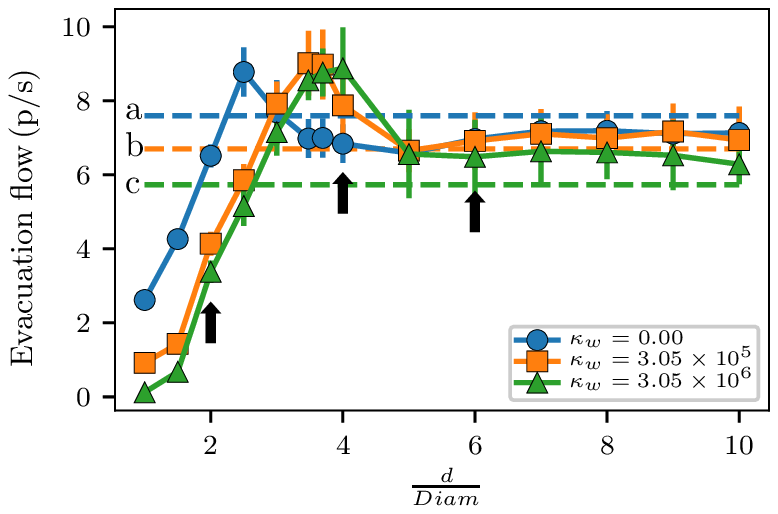}\label{flow_vs_d}}\ 
\subfloat[]{\includegraphics[width=0.49\columnwidth]
{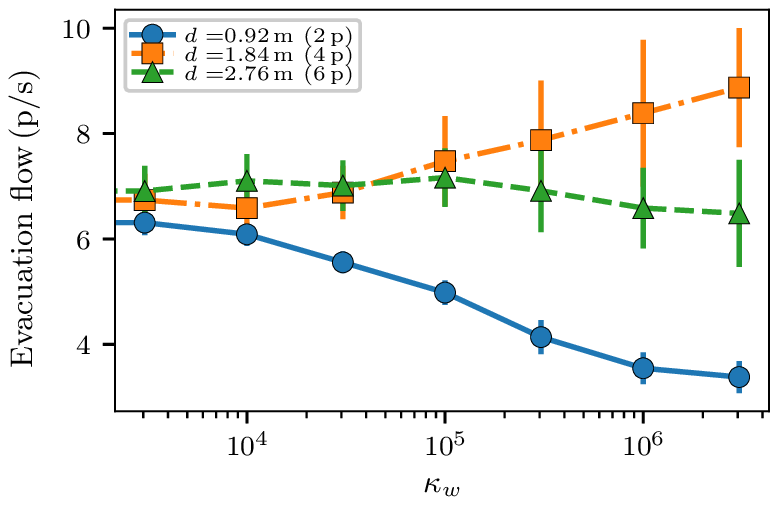}\label{flow_vs_kw}}\\
\caption[width=0.49\columnwidth]{ \textbf{(a)} Evacuation flow as a function 
of the distance between the panel-like obstacle and the exit door. The black 
arrows indicate the selected distances that were explored in 
Fig.\ref{flow_vs_kw}. The 
horizontal axis is scaled by the pedestrian's diameter $Diam=0.46\,$m. 
\textbf{(b)} Evacuation flow as a function of the wall friction 
coefficient. Data was averaged over 30 evacuation processes where 
the initial positions and velocities were random. The initial 
number of  pedestrians was $N=200$. The desired velocity was $v_d=6\,$m/s.}
\label{flow_vs_d_kw_all}
\end{figure}

This section introduces the effects of the wall friction and the distance from 
the panels to the exit. Fig.~\ref{flow_vs_d} shows the evacuation flow as a 
function of the distance $d$ between the panel and the exit door. Notice that 
the pedestrian's diameter scales the horizontal axis and each curve is 
associated with a particular friction coefficient value $\kappa_w$.\\

The horizontal dashed lines stand for the evacuation flow without an obstacle.  
Each line corresponds to a different $\kappa_w$ value; the letters `a', `b' and 
`c' in Fig.~\ref{flow_vs_d} stand for $\kappa_w=0$, and $3.05 \times 10^5$, 
$3.05 \times 10^6$ respectively. The flow values associated to these $\kappa_w$ 
are $J_a= (7.6\pm 0.7)\,$p/s, $J_b =(6.7 \pm 0.5)\,$p/s  and $J_c= (5.7 \pm 
0.5)\,$p/s. When the curves exceed the corresponding horizontal dashed lines, it 
means that placing the obstacle improves the evacuation performance.\\

The three curves exhibit similar qualitative behavior: for the lowest $d$ 
values, the curves increase until reaching a maximum, then, they decrease 
converging to a plateau. A similar behavior was reported in 
Ref.~\cite{echeverria2020pedestrian}, despite that pillar-like obstacles and 
spherocylinders agents were used.\\

Notice that each curve from Fig.~\ref{flow_vs_d} accomplishes a specific 
interval of $d$ in which placing an obstacle produces higher evacuation flows 
with respect to an obstacle-free room. In other words, placing a panel-like 
obstacle can improve or worsen the evacuation performance depending on its 
precise location with respect to the exit door. For example, if $\kappa_w=3.05 
\times 10^6$ the evacuation performance is improved only for $d/Diam > 3$ (see 
the triangle marker curve and the `c' horizontal line).  \\

The three curves show a maximum flow value for different values of $d$, while 
increasing the friction shifts the maximum value of the flow to higher values of 
$d$ (see Fig.~\ref{flow_vs_d}). This implies that the friction coefficient 
$\kappa_w$ has a significant influence on the evacuation. To further inspect 
this phenomenon, we show in Fig.~\ref{flow_vs_kw} the evacuation flow as a 
function of $\kappa_w$ for three representative values of $d$. These values are: 
$d=0.92\,$m, $1.84\,$m, $2.76\,$m.\\

If the vestibule is narrow ($d=0.92\,$m), increasing $\kappa_w$ reduces the 
evacuation flow because of the friction in the blocking clusters (see Section 
\ref{bc_results}). For a wider vestibule ($d=1.84\,$m), increasing $\kappa_w$ 
increases the flow because it prevents more pedestrians from accessing the 
vestibule (this effect is explained in detail in Section \ref{bc_results}). If 
the vestibule is too wide ($d=2.76\,$m), the wall frictions' role is almost 
negligible. This makes the flow remain roughly constant for all the $\kappa_w$ 
explored values.\\

These novel results suggests that the dynamics of the evacuations may be 
strongly influenced by the surface of the walls and obstacles. This fact 
was previously hypothesized by Hoogendoorn and 
Daamen~\cite{hoogendoorn2005pedestrian} but has not been tested numerically 
until the present research (to our knowledge).\\

\begin{figure}[!htbp]
\centering
\subfloat[]{\includegraphics[width=0.49\columnwidth]
{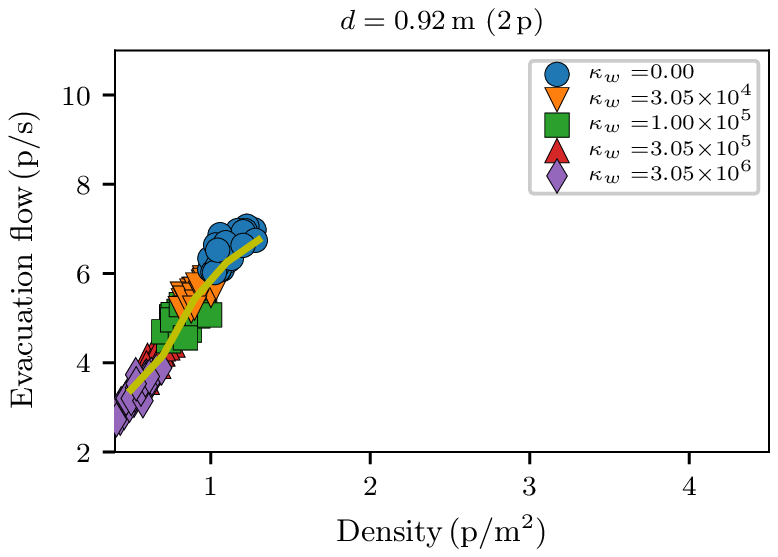}\label{flow_vs_density_d0_92}}\ 
\subfloat[]{\includegraphics[width=0.49\columnwidth]
{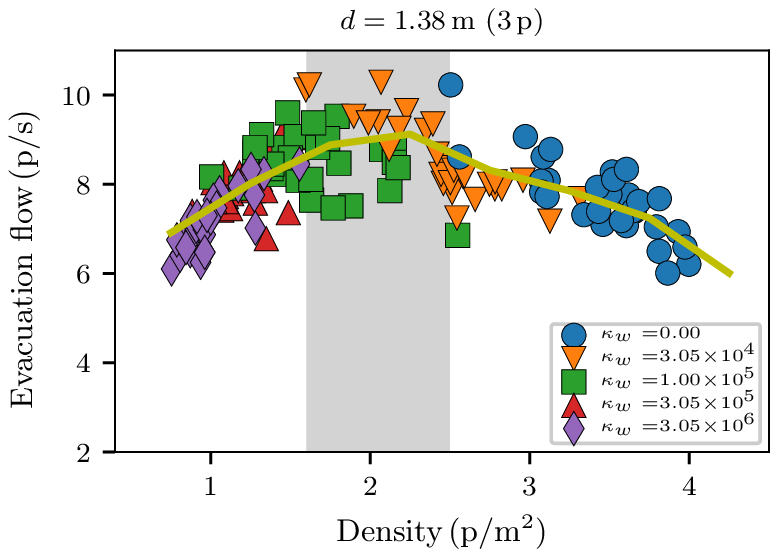}\label{flow_vs_density_d1_38}}\\
\subfloat[]{\includegraphics[width=0.49\columnwidth]
{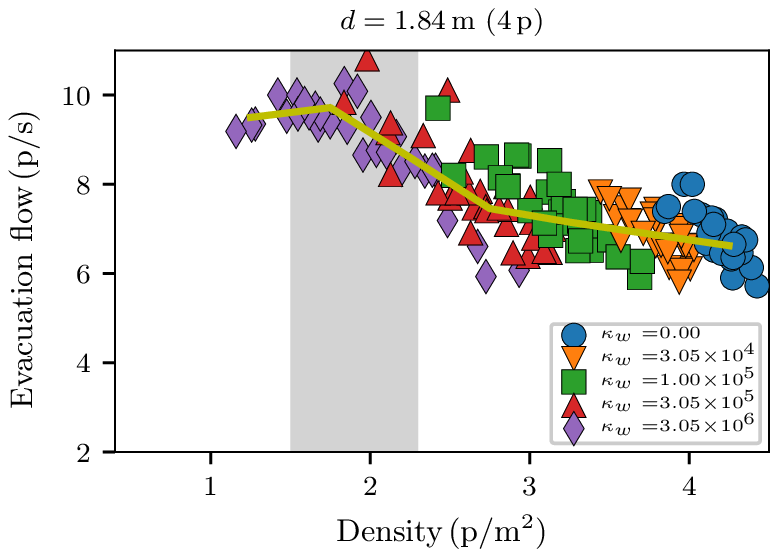}\label{flow_vs_density_d1_84}}\ 
\subfloat[]{\includegraphics[width=0.49\columnwidth]
{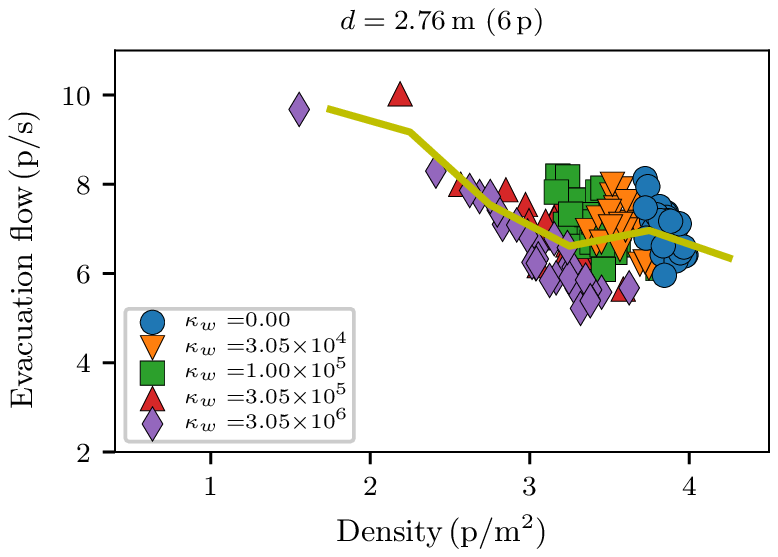}\label{flow_vs_density_d2_76}}\\
\caption[width=0.47\columnwidth]{Evacuation flow as a function of the density 
(fundamental diagram) for a two-entry vestibule. The 
density was measured on the inner vestibule and it was averaged over the 
evacuation time. There are 30 data points for each friction value. Each data 
point belongs to a different evacuation process where the initial condition 
(positions and velocities) were random. The initial number of pedestrians 
was $N=200$. The desired velocity was $v_d=6\,$m/s. See the plot's title for 
the corresponding distance from the obstacle to the door. The shady area 
highlights the densities that produce the maximum flow. }
\label{fundamental_diagram_1panel}
 \end{figure}

To further understand the mechanism behind this phenomenon, we studied 
the flow vs. density relation (the fundamental diagram). The density was 
sampled on the inner vestibule and it was averaged over the evacuation time. 
The results are shown in Fig.~\ref{fundamental_diagram_1panel} where each 
``subplot'' is associated with a fixed value of $d$ (see the plots' titles). The 
figures are ``scatter plots'' where each data point belongs to a single 
evacuation process 
(with random initial conditions). The colors stand for the friction 
coefficient value (see the legends). The solid yellow line appearing on each 
plot is the average of the scattered data points.\\

Before focusing on each subplot in Fig.~\ref{fundamental_diagram_1panel}, we 
want to mention two common features common on them.  The first salient 
feature is the negative correlation between the friction coefficient and the 
density (\textit{i.e.} the higher the friction, the lower the density values; 
check the colors of the data points). The second feature concerns the two 
regimes appearing in the fundamental diagram:  the 
free-flow regime (where increasing the density increases the flow) and the 
congested regime (where increasing the density reduces the flow). The situations 
explored in Fig.~\ref{fundamental_diagram_1panel} display either one or both 
regimes depending on the value of $d$.\\

We now proceed to the analysis of each subplot. Fig.~\ref{flow_vs_density_d0_92} 
corresponds to $d=0.92\,$m. Only the free-flow regime is 
observed for any of the $\kappa_w$ values explored. In this 
case, the flow does not attain the maximum because for such low-density 
values ($\rho < 2\,$p/m$^{2}$) there is left unused space in the vestibule. This 
means that more agents could potentially ingress the vestibule without 
producing congestion at the exit door.\\

The obstacle is then placed at $d=1.38\,$m in Fig.~\ref{flow_vs_density_d1_38}. 
This situation makes it possible to 
observe both regimes of the fundamental diagram (the free-flow and the 
congested regime). We also highlight (see the shady area) that the maximum flow 
values correspond to intermediate densities (say, $\rho \sim 2\,$p/m$^{2}$).\\

Figs.~\ref{flow_vs_density_d1_84} and \ref{flow_vs_density_d2_76} stand for 
$d=1.84\,$m and $d=2.76\,$m, respectively. 
Under these conditions, only the congested regime is observed. In other words, 
the entries to the vestibule are so large that the agents have almost no 
impediment to access the vestibule, which ultimately congests the area close to 
the door.\\

The four plots exhibit a clear relationship between the density and the flow. 
If the density is very low (say, $\rho<1.5\,$p/m$^{2}$) or very high (say, 
$\rho>3\,$p/m$^{2}$), the flow will not attain its maximum value. On the other 
hand, intermediate values for the density (say, $\rho \sim 2\,$p/m$^{2}$) 
maximize the evacuation flow because there is neither congestion nor 
left unused space in the vestibule; see the shady areas from 
Fig.~\ref{flow_vs_density_d1_38} and \ref{flow_vs_density_d1_84}.\\

The results shown in Fig.~\ref{fundamental_diagram_1panel} indicate that too 
low density (or too high) lead to a suboptimal situation with a somewhat 
reduced flow. Therefore, we claim that the best 
evacuation performance is achieved with intermediate density values. This is not 
in complete agreement with Refs.~\cite{zhao2017optimal,zhao2020experimental} 
where the authors suggest only a density reduction in the area close to the 
door. Thus, we suggest taking care of the obstacle distance and friction, in 
order to keep the vestibule density under control.\\

We emphasize that the density can either be controlled with $\kappa_w$ and $d$, 
while  the vestibule density determines the evacuation flow. Consequently, 
the evacuation flow can be ultimately optimized by adjusting $\kappa_w$ and $d$ 
for a given value of $v_d$.\\

Other researchers argue that the speed (hence the flow) is not uniquely 
determined by the density~\cite{garcimartin2017pedestrian}; thus, they introduce 
the kinetic stress observable in order to complete the description of the 
evacuation performance. In our research, however, it is unnecessary to 
include additional observables since the density measured on the inner vestibule 
is enough to determine the evacuation flow, at least in the context of the 
social force model.\\

\begin{figure}[!htbp]
\centering
{\includegraphics[width=0.7\columnwidth]
{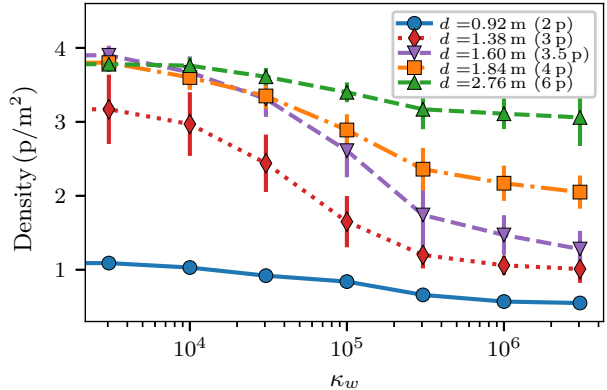}}\
\caption[width=0.49\columnwidth]{Density as a function of the friction 
coefficient. The density was measured on the inner vestibule and, it was 
averaged over the evacuation time until 90\% of individuals evacuated. The 
measurements were averaged over 30 evacuation processes, while the 
initial conditions (position and velocity) were random. The initial number 
of pedestrians was $N=200$. The desired velocity was $v_d=6\,$m/s. See the 
plot's legend for the corresponding distance from the obstacle to the door.}
\label{density_vs_kw_multi_d_door}
\end{figure}

As it was mentioned above that, for a given $v_d$, the parameters $\kappa_w$ 
and $d$ 
control the density in the inner vestibule. This phenomenon is more clearly 
depicted in Fig.~\ref{density_vs_kw_multi_d_door} where we show the density as a 
function of $\kappa_w$ for different $d$ values. Increasing $\kappa_w$ 
diminishes the 
density because pedestrians get stuck in the zones previous to the 
vestibule, which alleviates the congestion close to the exit door. This 
phenomenon will be further explained in the following Section.\\

In the same way, we observe in Fig.~\ref{density_vs_kw_multi_d_door} that 
reducing the distance between the obstacle and the door diminishes the density 
too. This phenomenon occurs because the vestibule entrance is smaller (which 
makes it more difficult for the pedestrians to access the vestibule). Thus, it 
is possible to reduce the density close to the exit door by either increasing 
$\kappa_w$ or reducing $d$ since both effects make it more difficult for 
pedestrians to enter the vestibule. \\

Although the results shown in 
Fig.~\ref{density_vs_kw_multi_d_door} correspond to a non-stationary situation, 
where the number of agents decreases over time, we show in \ref{appendix2} that 
the same results can be achieved in a stationary regime. \\
 
\subsubsection{\label{bc_results}The role of the blocking clusters}

We previously showed that the density of the inner vestibule can be 
regulated by $\kappa_w$ and $d$, while the evacuation flow depends 
on the density of the inner vestibule for a given $v_d$. The remaining question 
is, how do 
$\kappa_w$ and $d$ control the density. To answer this question, we studied the 
blocking clusters formed at the vestibule's entrances. Remember that the 
blocking clusters are clusters of pedestrians that block a door (in this 
case, we consider the doors as the two entries to the vestibule). See the 
Section \ref{bc} for a formal definition of blocking cluster. \\

\begin{figure}[!htbp]
\centering
\subfloat[]{\includegraphics[width=0.49\columnwidth]
{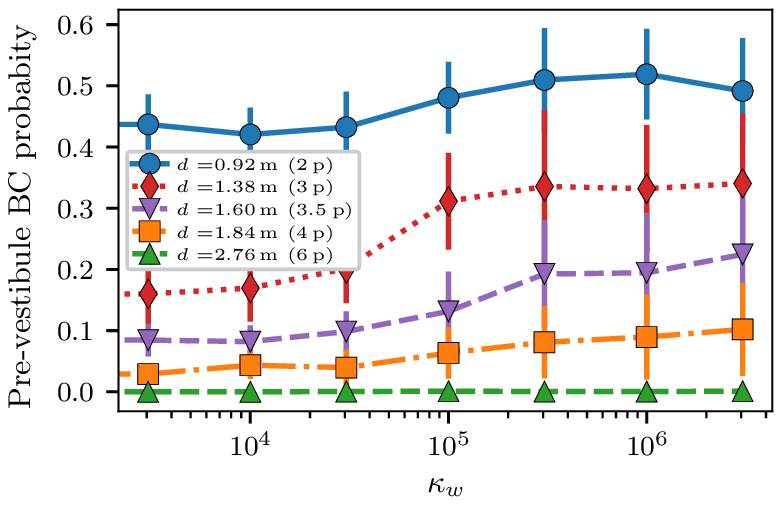}\label{bc_prevestibule_vs_kw_multi}}\ 
\subfloat[]{\includegraphics[width=0.49\columnwidth]
{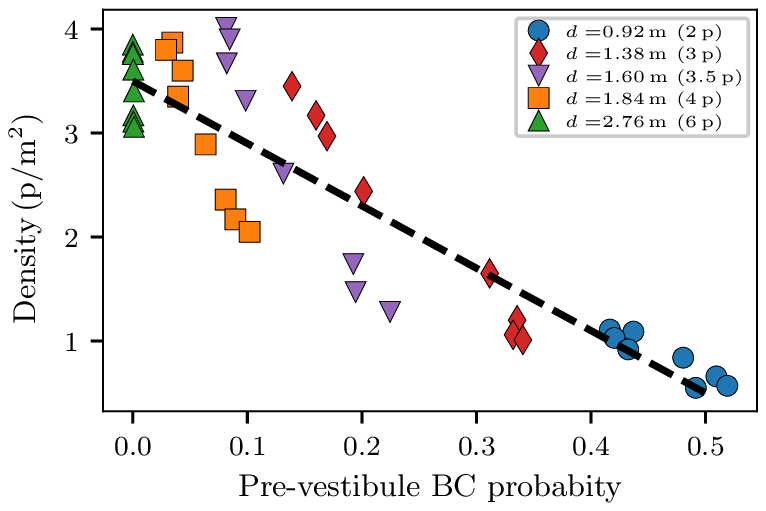}\label{bc_prevestibule_vs_density}}\\
\caption[width=0.49\columnwidth]{\textbf{(a)} Pre-vestibule blocking clusters 
probability as a function of the wall friction coefficient. The pre-vestibule 
blocking clusters is the mean value of the blocking cluster's probability taking 
into account the two entries to the vestibule. \textbf{(b)} Inner 
vestibule density as a function of the pre-vestibule blocking cluster 
probability. Each point corresponds to a different friction value. The dashed 
line is a visual guide to see the decreasing trend. The measurements were 
averaged over 30 evacuation processes for random initial conditions 
(position and velocity). 
The initial number of pedestrians was $N=200$. The desired velocity was 
$v_d=6\,$m/s. See the plot's 
legends for the corresponding distance of the obstacle to the door. }
\label{bc_prevestibule}
\end{figure}

Fig.~\ref{bc_prevestibule_vs_kw_multi} shows the probability of attaining a 
blocking cluster in the pre-vestibule zone as a function of $\kappa_w$ for 
different $d$ values. The pre-vestibule zone is the area before the entrance 
to the vestibule. Recall that the probability was defined as the fraction of 
time attaining blocking clusters over the total evacuation time. In 
Fig.~\ref{bc_prevestibule_vs_kw_multi} we report the mean value of the blocking 
cluster probability considering both entries to the vestibule.\\

Increasing the wall friction coefficient produces more persistent blocking 
clusters. This phenomenon can be noticed from the monotonically increasing 
behavior of the curves from Fig.~\ref{bc_prevestibule_vs_kw_multi}. In the same 
sense, the narrower the entries to the vestibule, the higher the blocking 
cluster probability. This effect is also carefully analyzed 
in Ref.~\cite{parisi2005microscopic} for the regular bottleneck case. \\

In summary, we found that increasing $\kappa_w$ or 
reducing $d$ produces an increment in the blocking cluster probability. The 
blocking cluster probability has a notorious impact on the vestibule density 
because it prevents pedestrians from entering the vestibule. The more blocking 
clusters present at the entrance of the vestibule, the less density the 
vestibule has, as shown in Fig.~\ref{bc_prevestibule}. In other words, the 
blocking clusters before the vestibule affect 
the density, and hence, the outgoing flow.\\

We may summarize our results from this Section as follows.  We found 
that the wall friction coefficient $\kappa_w$ and the distance between the 
obstacle and the door $d$ affect the vestibule density. The mechanism behind 
this involves the ``pre-vestibule'' blocking clusters (say, slowing 
down the access to the vestibule). Additionally, we showed that the 
inner vestibule density has a well established relationship with the evacuation 
flow 
(expressed through the fundamental diagram). The maximum flow occurs for 
intermediate values of density ($\rho \sim$ 2 p/m$^{2}$). We stress that if the 
density is very high or very low, the evacuation flow is suboptimal; but it was 
found that the presence of the vestibule clearly enhances the evacuation for 
rather high values of $v_d$ which can be related to stressful situations 
in which pedestrians can suffer considerable damage.\\

\subsection{\label{three-entry vestibule}The three-entry vestibule}

The previous Section dealt with evacuations in the presence of two-entry 
vestibule structures. The two-entry vestibule was designed by placing a single 
panel-like obstacle in front of the door. As a first step towards the analysis 
of more complex structures, we studied the evacuations through a three-entry 
vestibule structure designed from two panel-like 
obstacles (as illustrated in Fig.~\ref{three-entry_vestibule}). The desired 
velocity of the agents was set to $v_d=6\,$m/s.\\

The previous results showed that the vestibule's density (hence the 
evacuation flow) can be controlled by $\kappa_w$ and $d$. In this sense, 
the gap between the two panels introduces a new parameter capable of 
regulating the density. Notice that the gap works as a 
``middle entry'' to the vestibule. We intentionally choose a highly symmetric 
configuration in order to avoid path 
differences~\cite{escobar2003architectural}. \\

\begin{figure}[!htbp]
\centering
\subfloat[]{\includegraphics[width=0.49\columnwidth]
{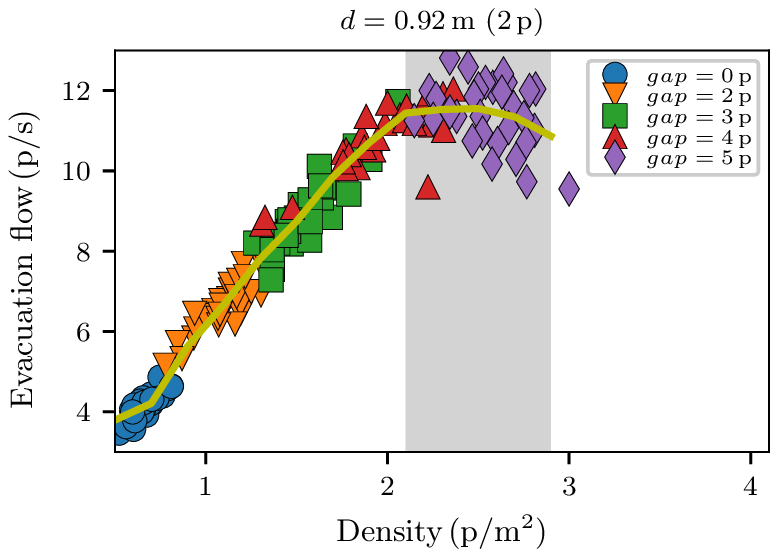}\label{flow_vs_density_2panels_d0_92}}\ 
\subfloat[]{\includegraphics[width=0.49\columnwidth]
{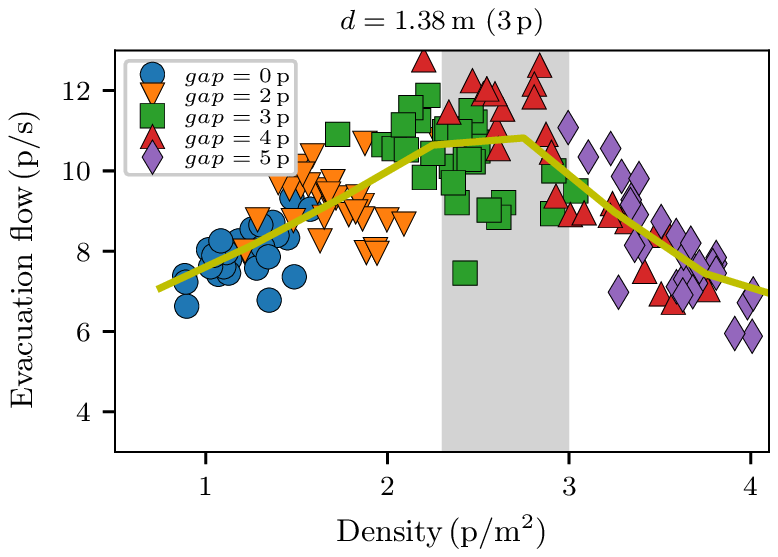}\label{flow_vs_density_2panels_d1_38}}\\
\subfloat[]{\includegraphics[width=0.49\columnwidth]
{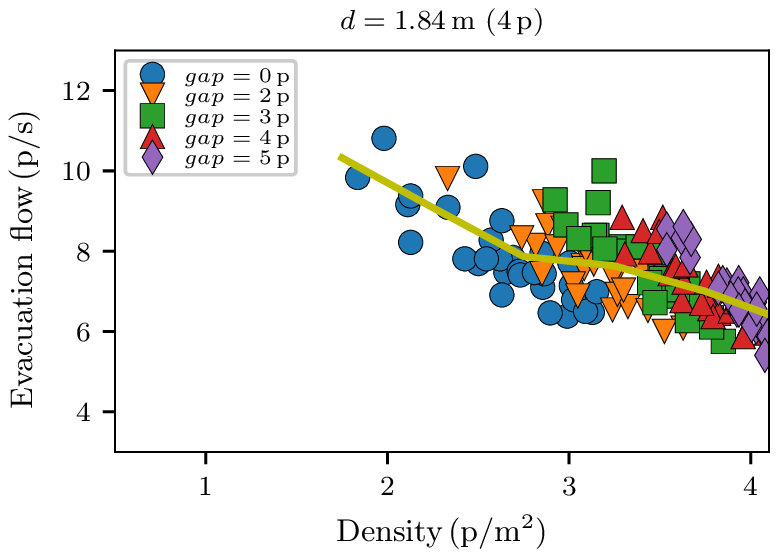}\label{flow_vs_density_2panels_d1_84}}\ 

\caption[width=0.47\columnwidth]{Evacuation flow as a function of the density 
(fundamental diagram) for a three-entry vestibule (see Fig.~\ref{rooms_layout} 
for the scheme). The density was sampled on 
the inner vestibule and it was averaged over the evacuation time. There are 30 
data points for each gap value. Each data point corresponds to a different 
evacuation process with random initial conditions. The initial number of 
pedestrians was $N=200$. The yellow line means the average over the data 
points. 
The desired velocity was $v_d=6\,$m/s. See the plot's 
title for the corresponding distance from the obstacle to the door. The 
shady area highlights the densities that produce the maximum flow. }
\label{fundamental_diagram_2panels}
\end{figure}

In this part of the investigation, we fix $\kappa_w$ and $v_d$ to focus only on 
the effect of the $gap$ between panels and the distance between the panels and 
the exit door $d$. Fig.~\ref{fundamental_diagram_2panels} shows fundamental 
diagrams (flow vs. density) for three representative $d$ values 
($d=0.92\,$m, $1.38\,$m, $1.84\,$m), where each plot is associated with a fix 
value of $d$ (see the title of the plots). We calculated the density and the 
evacuation flow following the same criteria as in the previous Section. Each dot 
in the plots corresponds to a single evacuation process (with random initial 
conditions).\\ 

Fig.~\ref{flow_vs_density_2panels_d0_92} displays 
the results for $d=0.92\,$m, the marker colors stand for the different $gap$ 
between panels. Notice that distinct values of the gap produce well 
distinguishable density intervals. For example, $gap=0\,$p produces densities 
below $\rho < 1\,$p/m$^{2}$, while $gap=5\,$p produces densities within 
the interval $2.5\,$p/m$^{2}< \rho \leq 3.0\,$p/m$^{2}$. It is worth mentioning 
that Fig.~\ref{flow_vs_density_2panels_d0_92} only attains the free-flow 
regime of the fundamental diagram (say, the higher the density, the higher the 
flow).\\

Fig.~\ref{flow_vs_density_2panels_d1_38} stands for $d=1.38\,$m. Now, the two 
regimes of the fundamental diagram are present (the free-flow and the congested 
regime) depending on the $gap$ value. If $gap=5\,$p, only the congested regime 
is observed, meaning that the agents can easily overcrowd the vestibule. Notice 
that placing the obstacles at $d=1.38\,$m produces more 
scattered results than placing the obstacle at $d=0.92\,$m. Nevertheless, it is 
still possible to 
associate different $gap$ values with well-distinguishable density values.\\

If the distance between panels and the exit door is $d=1.84\,$m 
(Fig.~\ref{flow_vs_density_2panels_d1_84}), the optimal evacuation flow can 
never be achieved because the density is too high, regardless of the gap value. 
This phenomenon occurs because the left and right entries to the vestibule are 
large enough to simultaneously facilitate the access of many agents, leading 
to 
congestion in the vestibule structure. Under this condition, the data points 
are even 
more scattered than in the previous results, making it more difficult to 
relate univocally the gap values with the vestibule density. \\

The results corresponding to the three-entry vestibule share some similarities 
with the results of the two-entry vestibule. In both cases, the optimal flow is 
attained for intermediate density values ($\rho \sim 2\,$p/m$^{2}$); compare, 
for instance, the shaded areas from Figs.~\ref{flow_vs_density_d1_38}, 
\ref{flow_vs_density_d1_84} and \ref{flow_vs_density_2panels_d0_92}, 
\ref{flow_vs_density_2panels_d1_38}. If the density is too high, in all the 
cases, it leads to a congested regime producing a suboptimal evacuation flow. If 
the density is too low, the flow is also below the optimal condition.\\

Beyond the similarities, it should be noted that the three-entry vestibule 
yields a maximum flow ($J \sim 11\,$p/s) that is higher than the maximum flow 
attained with the two-entry vestibule ($J \sim 8\,$p/s), see Tables 
\ref{table_two_entry} and \ref{table_three_entry} for details. This can be 
explained as follows, the two entries yield a suboptimal evacuation flow 
because there is space left inside the vestibule. The entry in the middle, 
however, increases the density driving the system to the optimal flow. This 
phenomenon could also be obtained with multiple entries.\\ 

Besides examining the evacuation flow, we briefly explored the 
pressure actuating on pedestrians during the evacuation process. We 
calculated the fraction of agents that withstand fatal pressures according to 
the criteria introduced in Ref.~\cite{cornes2017high}. We found that the 
vestibule reduces the fraction of agents exposed to dangerous pressures. This 
promising effect is more noticeable in three-entry vestibules. Further research 
should be conducted to analyze the role of hazardous pressures in evacuations 
with vestibule structures.\\

The final conclusion from this Section is that the gap between two panels 
can be used as an effective way to regulate the density in the vestibule and 
therefore maximize the evacuation flow. This result promotes creating novel 
architectural designs that can induce better evacuation performance in a panic 
situation. We stress, however, that we are in 
no way suggesting replacing the crowd management personnel or signals systems 
but instead examine the results shown in this paper as an architectural 
improvement for pedestrian evacuations.\\

\section{Conclusions}

Placing a panel-like obstacle in front of an exit door creates a vestibule 
structure that enhances the evacuation performance for intermediate 
door-obstacle distances ($d\sim3\,$p). Our investigation examined the effects 
of placing a two-entry vestibule (designed with only one panel-like obstacle) 
and a three-entry vestibule (designed with two aligned panel-like obstacles). \\

In both cases, we find that the density inside the vestibule uniquely determines 
the evacuation flow. If the vestibule gets jammed, the 
flow is diminished. If the density is too low, the flow is also in a suboptimal 
situation since there is unused space left in the vestibule. That is, more 
pedestrians could potentially enter the vestibule without producing congestion 
at the exit door. On the other hand, intermediate density values ($\rho \sim 
2\,$p/m$^{2}$) maximize the evacuation flow. This result 
holds for all the explored conditions in this work.\\

Our main conclusion is that at least three parameters are useful for regulating 
the density inside the vestibule (and therefore the evacuation flow). These 
parameters are the wall friction coefficient $\kappa_w$, the distance between 
the panel and the exit door $d$ and the gap between the two panels (in the case 
of a three-entry vestibule). It is worth mentioning that these parameters are 
architectural features that can be easily managed by designers and 
contractors.\\

Reducing $d$ or increasing $\kappa_w$ diminishes the density on the vestibule. 
This phenomenon is explained by means of the blocking clusters that appear at 
the vestibule entries. These blocking clusters limit the regular access to the 
vestibule. Likewise, reducing the gap between the two panels diminishes the 
density too.\\

We emphasize that the three-entry vestibule yields a maximum 
flow ($J \sim 11\,$p/s) that is higher than the maximum flow attained with the 
two-entry vestibule ($J \sim 8\,$p/s) and even much higher than the flow 
produced by a regular bottleneck without a vestibule ($J \sim 7\,$p/s). This 
result suggests that the three-entry vestibule structure could considerably 
improve emergency evacuations. \\

As a final remark, we would like to mention that in real-life emergency 
evacuations, individuals may not adopt a polite behavior in order to 
evacuate the building. It is therefore essential to design 
buildings that are ``panic-proof''. This means, buildings that could facilitate 
good evacuations even in the worst-case scenario where all the pedestrians try 
to escape selfishly. In this sense, we believe that finding the optimal 
vestibule structure is a step forward in the search for better human 
evacuations.\\

\section*{Acknowledgments}

This work was supported by the National Scientific and Technical 
Research Council (spanish: Consejo Nacional de Investigaciones Cient\'\i ficas 
y T\'ecnicas - CONICET, Argentina) grant Programaci\'on Cient\'\i fica 2019 
(UBACYT) Number 2019-2019-01994.\\

G.A Frank thanks Universidad Tecnol\'ogica Nacional (UTN) for partial
support through Grant PID Number SIUTNBA0006595.\\

\appendix

\section{\label{appendix1}Flow-density quantitative results}

In this appendix, we provide a summary of the numerical results for the flow and 
the density. All the results are presented in two tables that exhibit the mean 
values and standard deviations. Tab. \ref{table_two_entry} corresponds to the 
two-entry vestibule while Tab. \ref{table_three_entry} corresponds to the 
three-entry vestibule. \\

\begin{table}[H]
\centering
\vspace*{0.1cm} 
\textit{\hspace*{1cm} Two-entry vestibule}
\begin{tabular}{c@{\hspace{6mm}}c@{\hspace{6mm}}c@{\hspace{
6mm} } c@ {\hspace{6mm}} c@{\hspace{6mm}} c@{\hspace{14mm}}l}
\toprule
 $d\,$(m) & $\kappa_w$  & $\langle J \rangle \,$(p/s)    & $\sigma_J\,$(p/s)   
&  $\langle \rho \rangle \,$(p/m$^{2}$) & $\sigma_{\rho}\,$(p/m$^{2}$) \\
\toprule
0.92  &  0 & 6.5 & 0.3 & 1.10 & 0.08  \\
0.92  & $3.05\times10^4$ & 5.5 & 0.2 & 0.92 & 0.06  \\
0.92  & $1.00\times10^5$ & 4.9 & 0.2 & 0.83 & 0.06 \\
0.92  & $3.05\times10^5$ & 4.1 & 0.3 & 0.66 & 0.07 \\
0.92  & $3.05\times10^6$ & 3.3 & 0.3 & 0.54 & 0.06 \\
\hline
1.38  &  0 & 7.6 & 0.8 & 3.44 & 0.38  \\
1.38  & $3.05\times10^4$ & 8.5 & 0.8 & 2.44 & 0.39 \\
1.38  & $1.00\times10^5$ & 8.4 & 0.6 & 1.64 & 0.35 \\
1.38  & $3.05\times10^5$ & 7.9 & 0.6 & 1.19 & 0.18 \\
1.38  & $3.05\times10^6$ & 7.1 & 0.6 & 1.00 & 0.19 \\
\hline
1.84  &  0 & 6.8 & 0.5 & 4.18 & 0.14  \\
1.84  &  $3.05\times10^4$ & 6.8 & 0.5 & 3.79 & 0.16  \\
1.84  & $1.00\times10^5$ & 7.4 & 0.8 & 3.16 & 0.30 \\
1.84  & $3.05\times10^5$ & 7.8 & 1.1 & 2.64 & 0.35 \\
1.84  & $3.05\times10^6$ & 8.8 & 1.1 & 1.93 & 0.45 \\
\hline
2.76  &  0 & 6.9 & 0.4 & 3.84 & 0.07  \\
2.76  &  $3.05\times10^4$ & 7.0 & 0.4 & 3.60 & 0.12  \\
2.76  & $1.00\times10^5$ & 7.1 & 0.5 & 3.39 & 0.13 \\
2.76  & $3.05\times10^5$ & 6.9 & 0.7 & 3.17 & 0.27 \\
2.76  & $3.05\times10^6$ & 6.4 & 1.0 & 3.05 & 0.39 \\
\hline \end{tabular}
\caption{Mean value and standard deviation of the flow ($J$) and the 
density ($\rho$) for different $d$ and $\kappa_w$. The desired velocity was 
$v_d=6\,$m/s.}
\label{table_two_entry}
\end{table}

\begin{table}[H]
\centering
\vspace*{0.1cm} 
\textit{\hspace*{1cm} Three-entry vestibule}
\begin{tabular}{c@{\hspace{6mm}}c@{\hspace{6mm}}c@{\hspace{6mm}}c@
{\hspace{6mm}}
c@{\hspace{6mm}}
c@{\hspace{14mm}}l}
\toprule
 $d\,$(m) & $gap\,$(p)  & $\langle J \rangle \,$(p/s)    & $\sigma_J\,$(p/s)   
&  $\langle \rho \rangle \,$(p/m$^{2}$) & $\sigma_{\rho}\,$(p/m$^{2}$) \\
\toprule
0.92  & 0 & 4.1 & 0.3 & 0.66 & 0.07  \\
0.92  & 2 & 6.6 & 0.6 & 1.10 & 0.14  \\
0.92  & 3 & 8.9 & 0.9 & 1.56 & 0.17 \\
0.92  & 4 & 10.8 & 0.9 & 1.98 & 0.29 \\
0.92  & 5 & 11.4 & 0.8 & 2.56 & 0.20 \\
\hline
1.38  & 0 & 7.9 & 0.6 & 1.19 & 0.18  \\
1.38  & 2 & 9.2 & 0.7 & 1.70 & 0.23  \\
1.38  & 3 & 10.2 & 0.9 & 2.38 & 0.27 \\
1.38  & 4 & 10.0 & 1.9 & 2.99 & 0.43 \\
1.38  & 5 & 8.0 & 1.3 & 3.57 & 0.26 \\
\hline
1.84  & 0 & 7.8 & 1.1 & 2.64 & 0.35  \\
1.84  & 2 & 7.5 & 0.8 & 3.14 & 0.28  \\
1.84  & 3 & 7.5 & 0.9 & 3.44 & 0.24 \\
1.84  & 4 & 7.0 & 0.8 & 3.75 & 0.23 \\
1.84  & 5 & 6.9 & 0.8 & 3.91 & 0.18 \\
\hline \end{tabular} 
\caption{Mean value and standard deviation of the flow ($J$) and the 
density ($\rho$) for different $d$ and $gap$. The wall friction coefficient was 
$\kappa_w=3.05\times 10^5$ and the desired velocity $v_d=6\,$m/s. }
\label{table_three_entry}
\end{table}

\section{\label{appendix2}Density in the stationary regime}

The purpose of this appendix is to show that the density measurements in the 
non-stationary regime are equivalent to the measurements in the stationary 
regime. By ``stationary'' we mean that the number of agents is constant in 
time; we achieve this by re-entering the outgoing agents in order to keep 
$N=200$.\\

\begin{figure}[!htbp]
\centering
{\includegraphics[width=0.7\columnwidth]
{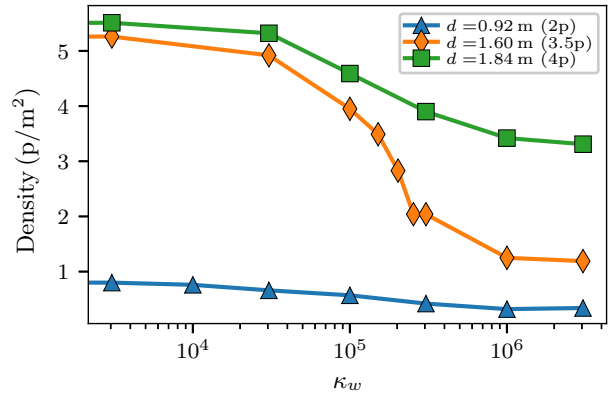}}\
\caption[width=0.49\columnwidth]{Density as a function of the friction 
coefficient. The density was measured on the inner vestibule. The 
measurements are averaged over $\Delta t=1000\,$s. The number of pedestrians was 
$ N = 200 $ all the time since the outgoing agents were re-entered. The 
desired velocity was $v_d=6\,$m/s. See the plot's legend for the corresponding 
distance from the obstacle to the door.}
\label{density_vs_kw_multi_d_door_stationaty}
\end{figure}

Fig.~\ref{density_vs_kw_multi_d_door_stationaty} shows the density in the inner 
vestibule as a function of the friction coefficient for different $d$ values. 
The measurements correspond to a stationary regime that lasts $\Delta 
t = 1000\,$s. The qualitative behavior from 
Fig.~\ref{density_vs_kw_multi_d_door_stationaty} is similar to the qualitative 
behavior from Fig.~\ref{density_vs_kw_multi_d_door} (which corresponds to a 
non-stationary situation).  This means that the density dependence on $\kappa_w$ 
and $d$ is not an effect produced by the decreasing number of pedestrians over 
time. \\

\section*{References}

\bibliography{bibfile}

\end{document}